\documentclass[12pt,psfig]{article} \textheight=23 true cm \textwidth=16 true cm
\usepackage{graphicx}
\usepackage{amsmath}
\usepackage{amssymb}

\begin{document}

\begin{center}
{\Large\bf Quintessential Phenomena in Higher Dimensional Space Time }\\[15 mm]
D. Panigrahi\footnote{Relativity and Cosmology Research Centre,
Jadavpur University, Kolkata - 700032, India , e-mail:
dibyendupanigrahi@yahoo.co.in , Permanent Address : Sree Chaitanya
College, Habra, 743268, India}
  and S. Chatterjee\footnote{Relativity and Cosmology Research Centre, Jadavpur University,
Kolkata - 700032, India, Permanent Address : IGNOU Convergence Centre, New Alipore College,
Kolkata - 700053, India,  e-mail : chat\_sujit1@yahoo.com\\Correspondence to : S. Chatterjee} \\[10mm]

\end{center}

\begin{abstract}
The higher dimensional cosmology provides a natural setting to
treat, at a classical level, the cosmological effects of vacuum
energy. Here we discuss two situations where starting with an
ordinary matter field without any equation of state we end up with
a Chaplygin type of gas apparently as a consequence of extra
dimensions. In the second case we study the quintessential
phenomena in higher dimensional spacetime with the help of a
Chaplygin type of matter field. The first case suffers from the
disqualification that no dimensional reduction occurs, which is,
however, rectified in the second case. Both the models show the
sought after feature of occurrence of \emph{flip} in the rate of
expansion. It is observed that with the increase of dimensions the
occurrence of \emph{flip} is delayed for both the models, more in
line with current observational demands. Interestingly we see that
depending on some initial conditions our model admits QCDM,
$\Lambda$CDM and also Phantom like evolution within a unified
framework. Our solutions are general in nature in the sense that
when the extra dimensions are switched off the known 4D model is
recovered. Correspondence to a recent work of Guo etal for a
\emph{quiessence} like model is also found.
\end{abstract}
\bigskip
\section*{1. Introduction}

Distance measurement of type Ia supernovae as well as cosmic
microwave background anisotropy measurements in the last decade
points to a late accelerated expansion of the universe.
Gravitational interaction being always attractive in nature this
finding, corroborated by a number of cosmic probes has been
puzzling astronomers for a long time and has so far evaded any
plausible but consistent explanation based on sound physical
principles. Briefly stated - proposals include among others flavor
oscillations of axions ~\cite{kalop}, modification of
Einstein-Hilbert action via the additional curvature terms in the
Lagrangian ~\cite{ua}, introduction of
 an evolving cosmological constant in the field equations, the
 role of inhomogeneity to cause acceleration ~\cite{aln}, Brans-Dicke scalar field
  and finally the presence of a quintessential type of scalar
   field giving rise to a large negative pressure(dark
   energy)~\cite{pd}.\vspace{0.5cm}

\hspace{-0.76 cm} In this context the authors of the present
article have been, of late, struggling
 with the idea of explaining the late acceleration as a higher
 dimensional(HD) effect ~\cite{dp, sc1}. In the framework of higher dimensional
 cosmology we have been able to show, though in a rather naive
 way, that the acceleration can be explained as a consequence of
 the presence of the extra spatial dimensions and this effect has
 been  coined as `\emph{dimension driven}' accelerating model.
 In fact here the effective Friedmann equations contain additional
 terms coming from the extra dimensions which may be viewed as a
 `fluid' responsible for the current acceleration. So here we
 attempt to incorporate the phenomenon of acceleration within the
 framework of HD spacetime itself without invoking a mysterious
 scalar field with large negative pressure by hand. On the contrary
 the origin of the extra fluid responsible for the acceleration is
 geometric in origin having strong physical foundation and more in
 line with the spirit of general relativity as proposed by Einstein~\cite{einstein}
 and later developed by Wesson and his collaborators\cite{wes}. Moreover in
 a recent communication~\cite{milton} it is argued that quantum fluctuation in
 4D spacetime do not give rise to dark energy but rather a
 possible source of the dark energy is the fluctuations in the
 quantum fields including quantum gravity inhabiting extra
 compactified dimensions. So we recently witness a spurt of
 activities relating to higher dimensional spacetime in its
 attempts to unify gravity with other forces of nature,
 development of varied brane models, induced matter proposal and
 very recently the dimension driven quintessential models.
 The present investigation is primarily motivated by two considerations.
 While we have plenty of multidimensional cosmological models in
 literature~\cite{bron} and also some sporadic works of Chaplygin
 inspired brane models ~\cite{mak, her} we are not much aware of models
 of similar kind in higher dimensions either driven by extra dimensions
 themselves or by Chaplygin type of gas.\vspace{0.5cm}

\hspace{-0.8 cm} The present work essentially comprises two parts.
We have taken a
 (d+4) dimensional homogeneous spacetime with two scale factors and a
 perfect fluid as a source field. Assuming an ansatz in the form
 of the deceleration parameter we find an expression of the scale
 factor for isotropic expansion. In the process we get a class of
 solutions for the matter field with interesting physical
 properties closely mimicing the well known expression of a
 generalised Chaplygin gas suitable for accelerating model. One
 may look upon the resulting matter field as a two-component
 fluid consisting a cosmological constant and a perfect fluid
 obeying a higher dimensional equation of state. One finds that in
 the early phase the pure matter field predominates  resulting in
 a decelerated expansion followed by the predominance of the
 cosmological constant with the expansion transiting to an
 accelerating phase, while in the intermediate stages our
 cosmology interpolates between the different phases of the
 universe. For sake of completeness we also construct an
 equivalent scalar field with its potential energy term to
 simulate the dynamics of the cosmology with the actual matter
 field.\vspace{0.5cm}

\hspace{-0.6 cm}In the second part of the program the approach is
 reversed. Here we start with the more conventional approach of
 taking a Chaplygin type of matter field to start with. Our first
 model suffers from the disqualification that both 4D and HD scale
 factors are expanding so the final reduction to the current
 manifestly 4D universe does not work. This shortcoming is
 remedied in the second part. While literature abounds with
 quintessential models with Chaplygin type of gas in 4D we are not
 aware of results of similar kind in HD spacetime. However we have
 not been  able to find solutions in a closed form with the
 system of equations finally reducing to a hypergeometric
 series. In any case certain inferences can always be drawn in the
 extreme cases and our analysis shows that an initially
 decelerating model transits to an accelerating one as in 4D. An
 interesting result in this section is the fact that the effective
 equation of state at the late stage of evolution contains some
 additional term coming from extra dimensions. Fixing some initial
 conditions the cosmology evolves as QCDM, $\Lambda$CDM or Phantom
 type. Though not exactly similar this points to the
 `\emph{k-essence}' type of models which lead to cosmic
 acceleration today for a wide range of initial conditions without
 fine-tuning and without invoking an anthropic
 argument.

\section*{2. Accelerating Universe - I}

We  consider the line element of (d+4)-dimensional spacetime
\begin{eqnarray}
  ds^{2} &=&
  dt^{2}-R^{2}\left(\frac{dr^{2}}{1-r^{2}}+r^{2}d\theta^{2}+
  r^{2}sin^{2}\theta d\phi^{2}\right) - A^{2}\gamma_{ab}dy^{a}dy^{b}
\end{eqnarray}
where $y^{a}$ ( $a,b = 4,....        , 3+d$) are the extra
dimensional spatial coordinates and the 3D and extra dimensional
scale factors $R $ and $A $ depend on time only  and $K$ is the 3D
curvature and the compact manifold is described by the metric
$\gamma_{ab}$. For our manifold $M^{1}\times S^{3}\times S^{d}$
the symmetry group of the spatial section is $O(4) \times O(d+1)
$. The stress tensor whose form will be dictated by Einstein's
equations must have the same invariance leading to the energy
momentum tensor as \cite{rd}
\begin{equation}
T_{00}=\rho~,~~T_{ij}= -p(t)g_{ij}~,~~ T_{ab}=-p_{d}(t)g_{ab}
\end{equation}
where the rest of the components vanish. Here $p$ is the isotropic
3-pressure and $p_{d}$, that in the extra dimensions.
\vspace{0.5cm}

\hspace{-0.8 cm} The independent field equations  for our metric
(1) are

\begin{eqnarray}
\rho &=& 3\frac{\dot{R^{2}}}{R^{2}}  +
\frac{1}{2}d(d-1)\frac{\dot{A^{2}}}{A^{2}}
+3d\frac{\dot{R}\dot{A}}{RA}
\end{eqnarray}
\begin{eqnarray}
 p &=&-2\frac{\ddot{R}}{R}-\frac{\dot{R^{2}}}{R^{2}}-
  - d \frac{\ddot{A}}{A} -
 \frac{1}{2}d(d-1)\frac{\dot{A^{2}}}{A^{2}}-2d\frac{\dot{R}}{R}\frac{\dot{A}}{A}
\end{eqnarray}
\begin{eqnarray}
p_{d} &=& - 3 \frac{\ddot{R}}{R} - 3\frac{\dot{R^{2}}}{R^{2}} -
 (d-1)\frac{\ddot{A}}{A} -
\frac{1}{2}(d-1)(d-2)\frac{\dot{A^{2}}}{A^{2}}-3(d-1)\frac{\dot{R}\dot{A}}{RA}
 \end{eqnarray}
 Here we have three independent equations with five unknowns $(\rho
 , p,p_{d},A,R)$ and hence we are at liberty to conveniently fix up two
  initial conditions. Firstly we take a perfect fluid with
  isotropic pressure in all dimensions such that $p = p_{d}$
  and secondly we assume a particular form of deceleration
  parameter as
\begin{equation}
q = \frac{a - R^{w}}{b + R^{w}}
\end{equation}
where $a, b$ and $w$ are arbitrary constants. The equation (6)
yields

 \begin{eqnarray}
 R(t) = R_{0} \mathrm{sinh}^{n}\omega t
 \end{eqnarray}($n= \frac{2}{w}$)
such that we get
\begin{equation}
q = \frac{1-n~\mathrm{cosh}^{2}\omega t}{n~\mathrm{cosh}^{2}\omega
t}
\end{equation}
showing that the exponent $n $ determines the evolution of $q$.
While for $ n > 1$, it is only accelerating but for $ n < 1 $ we
are able to achieve the desirable feature of \emph{flip}, although
it is not obvious from our analysis at what value of redshift this
\emph{flip} occurs.
 The second assumption $p = p_{d}$ gives
\begin{eqnarray}
- \frac{\ddot{R}}{R} - 2 \frac{\dot{R}^{2}}{R^{2}} +
\frac{\ddot{A}}{A} + (d-1)\frac{\dot{A}^{2}}{A^{2}} +
(3-d)\frac{\dot{R}\dot{A}}{RA} = 0
\end{eqnarray}
Relevant to point out that $R = A$ identically satisfies this
equation and in this case the matter density and isotropic
pressure come out to be

\begin{eqnarray}
\rho &=& \frac{1}{2}(d+2)(d+3)\frac{\dot{R^{2}}}{R^{2}}
\end{eqnarray}
\begin{eqnarray}
 p = p_{d} =-(d+2)\frac{\ddot{R}}{R}- \frac{1}{2}(d+1)(d+2)\frac{\dot{R^{2}}}{R^{2}}
 \end{eqnarray}
Now we are in a position to calculate the expressions for matter
density and pressure and find
\begin{eqnarray}
\rho =  \Lambda + \frac{B}{R^{\frac{2}{n}}}
\end{eqnarray}
\begin{eqnarray}
 p = p_{d} = - \Lambda - \frac{n(d+3)-2}{n(d+3)}\frac{B}{R^{\frac{2}{n}}}
\end{eqnarray} where, $\Lambda = \frac{1}{2}(d+2)(d+3)n^{2}\omega^{2}$ and $B =
\Lambda R_{0}^{\frac{2}{n}}$.\\
We also get the equation of state as
\begin{eqnarray}
 p = (\gamma -1) \rho -  \frac{2 \Lambda}{(d+3)n} = (\gamma - 1)
 \rho - \gamma \Lambda
\end{eqnarray}
 where, $(\gamma -1) = - \frac{(d+3)n - 2}{(d+3)n}$,  so $ n =
\frac{2}{\gamma (d+3)}$ and   $ \omega = \gamma
\sqrt{\frac{(d+3)\Lambda}{2(d+2)}}$ such that we finally get
\begin{eqnarray}
 R(t) = R_{0}\left[ \mathrm{sinh} \sqrt{\frac{(d+3)\Lambda}{2(d+2)}}\gamma t
 \right]^{\frac{2}{\gamma (d+3)}}
  \end{eqnarray}
  The last equation makes interesting reading. As $t \sim~ 0$ , $sinh
  t\sim t$ and we get
  \begin{equation}
 R = t^\frac{2}{\gamma (d + 3)}
  \end{equation}
  such that for 4D case($d = 0$)\cite{mnras}
\begin{equation}
R = t^{\frac{2}{3 \gamma}}
\end{equation}
which yields the well known solutions as $ t = t^{\frac{1}{2}}$, $
t = t^{\frac{2}{3}}$ for radiation and dust respectively for 4D.
 Replacing $n$ in the matter field equations we finally get
\begin{equation}
\rho =  \Lambda + \frac{B}{R^{\gamma (d+3)}}~~ \mathrm{and}~~ p =
p_{d} = - \Lambda -( 1-\gamma )\frac{B}{R^{\gamma (d +3)}}
\end{equation}
\vspace{0.5cm}

\hspace{-0.6 cm}From our analysis it follows that\vspace{0.5cm}

(\emph{i}) for dust, $\gamma = 1$ and $n = \frac{2}{d+3}$ \\
\begin{equation}
\rho = \Lambda +  \frac{B}{R^{(d+3)}} ~~~ \mathrm{and} ~~~ p = -
\Lambda  ~~~~~
\end{equation}

(\emph{ii}) for radiation, $\gamma = \frac{d+4}{d+3}$ and $n =
\frac{2}{d+4}$ \vspace{0.5cm}

\begin{equation}
\rho = \Lambda +  \frac{1}{R^{(d+4)}} ~~~ \mathrm{and} ~~~ p = -
\Lambda + \frac{1}{(d+3)}\frac{B}{R^{(d+4)}}
\end{equation}

(\emph{iii}) for stiff fluid, $\gamma = 2$ and $n = \frac{1}{d+3}$
\vspace{0.5cm}

\begin{equation}
\rho = \Lambda +  \frac{B}{R^{2(d+3)}} ~~~ \mathrm{and} ~~~ p = -
\Lambda + \frac{B}{R^{2(d+3)}}
\end{equation}\vspace{0.5cm}

\hspace{-0.5 cm}One can calculate $q$ for different values of
$\gamma$. Taking \vspace{0.5cm}

(\emph{i}) $\gamma= 2 $  (stiff fluid)
\begin{equation}
 q = \frac{(d+2) - \mathrm{sinh}^{2}\omega t}{1 + \mathrm{sinh}^{2} \omega t}
\end{equation}

(\emph{ii}) $\gamma = \frac{d +4}{d+3}$ ( radiation)
\begin{equation}
 q = \frac{(d+2) - 2 ~\mathrm{sinh}^{2}\omega t}{2 + 2~ \mathrm{sinh}^{2} \omega t}
\end{equation}

(\emph{iii}) $\gamma = 1$ (dust)
\begin{equation}
 q = \frac{(d+1) -2~ \mathrm{sinh}^{2}\omega t}{2 + 2 ~\mathrm{sinh}^{2} \omega t}
\end{equation}

\begin{figure}[h]

\begin{center}
  \includegraphics[width=10 cm]{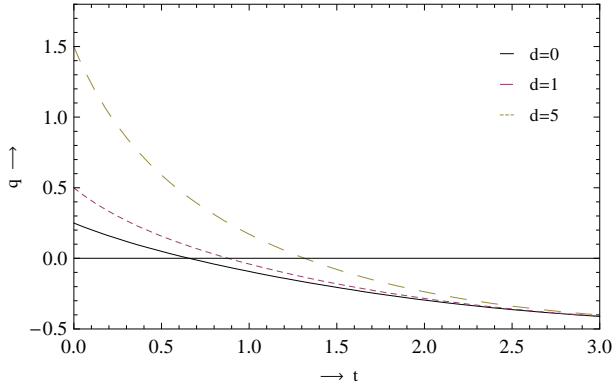}
  \caption{
  \small\emph{The variation of $q$ and $t$  for different
  values of $d$  in dust dominated era. Flip is delayed as
  dimensions increases.
   }\label{1}
    }
\end{center}
\end{figure}\vspace{0.5cm}

\hspace{-0.7 cm} At this stage one can check the influence of
extra dimension on the occurrence of \emph{flip}. The figure-1
shows that with dimensions the instant of \emph{flip} is delayed.
The fact that the time of \emph{flip} is delayed with dimensions
is in line with observational demands. This is because there is
plenty of observational evidence for a decelerated universe in the
recent past~\cite{ponce}. However the dominance of vacuum energy
over ordinary matter  at some $z_{eq}>0$ does not guarantee the
present acceleration of the universe. For this, the vacuum energy
has to dominate long enough as to overcome the gravitational
attraction produced by ordinary matter. Current dynamical mass
measurements suggest that the matter content of the universe adds
up to 30 percent of the critical density. It is generally believed
that the transition from deceleration to acceleration occurs at a
redshift $0.28 < z_{T}< 0.68$, which confirms the idea that the
acceleration is a recent phenomenon. Hence the significance of the
above plot is particularly encouraging in the context of higher
dimensional cosmology.\vspace{0.5cm}

\hspace{-0.8 cm} Now the expression of speed of sound will be
\begin{equation}
C_{s}^{2} = \frac{\delta p}{\delta \rho} = \frac{2}{\gamma} -1
\end{equation}
 \\which gives (\emph{i}) for dust case: $C_{s}^{2} = 0$; (\emph{ii}) for radiation
case : $C_{s}^{2} =\frac{ 1}{d+3}$ and (\emph{iii}) for stiff
fluid : $C_{s}^{2} = 1$. Thus we see that the velocity of sound
vanishes in the dust model as expected and it is maximum in 4D and
decreases with dimension in radiation dominated era. To avoid
imaginary value of the speed of the sound $\gamma < 2$. Again
$C_{s}$ should never exceed the speed of light for the range of
$\gamma$,   $1 < \gamma < 2$. \vspace{0.5cm}

\hspace{-0.75 cm} As is well known the whole dynamics can be
simulated in a field theoretic approach with the help of an
equivalent scalar field~\cite{barrow}. For a scalar field $\phi$
having self-interacting potential $U(\phi)$, the Lagranian of the
scalar field will be $\mathcal{L} =
\frac{1}{2}\partial^{\mu}\phi\partial_{\mu}\phi - U(\phi)$, the
equivalent energy density $\rho_{\phi}$ and pressure $p_{\phi}$
for the scalar field will be
\begin{equation}
\rho_{\phi} =\frac{1}{2}\dot{\phi^{2}} + U(\phi) = \Lambda +
\frac{B}{R^{\gamma (d+3)}}
\end{equation}
and the corresponding `pressure' as
\begin{equation}
p_{\phi} =\frac{1}{2}\dot{\phi^{2}} - U(\phi) =- \Lambda -
\frac{B}{R^{\gamma (d +3)}} + \gamma \frac{B}{R^{\gamma (d+ 3)}}
\end{equation}
such that
\begin{equation}
\dot{\phi^{2}} = \gamma \frac{B}{R^{\gamma (d+ 3)}}
\end{equation}
which, in turn, gives via equation (3) for (d+4) spacetime
\begin{equation}
\phi' = \pm \sqrt{\frac{B\gamma (d+2)(d+3)}{2}}\frac{1}{R
\sqrt{\Lambda R^{\gamma (d +3)}+B}}
\end{equation}
(here a prime and a dot overhead denote differentiation w.r.t. $R$
and $t$ respectively.)
\begin{equation}
\phi = \sqrt{\frac{2(d+2)}{\gamma (d+3)}}~
\mathrm{tanh}^{-1}\left(\sqrt{\frac{\Lambda R^{\gamma (d+3)}+
B}{B}}\right)
\end{equation}
or
\begin{equation}
\phi = \sqrt{\frac{2(d+2)}{\gamma (d+3)}}~
\mathrm{tanh}^{-1}\left(\mathrm{cosh} \omega t\right)
\end{equation}
 and
\begin{equation}
U( \phi ) = \frac{ \Lambda \gamma}{2}\left\{ 1+ \left( 1-
\frac{2}{\gamma} \right) \sinh^{2}\sqrt{\frac{1+
(1-\frac{2}{\gamma} (d+3)}{2(d+2)}}\phi\right\}
\end{equation}

\begin{figure}[h]

\begin{center}
  \includegraphics[width=10 cm]{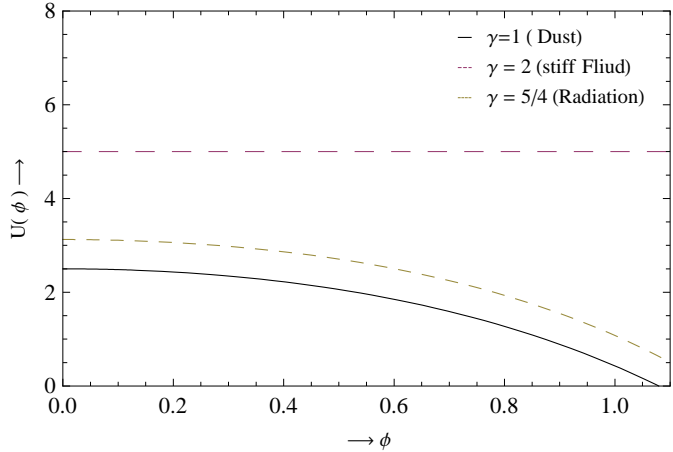}\\
  \caption{
  \small\emph{The variation of $U(\phi)$ and $\phi$ is shown in this figure for different values of
  $\gamma$ in 5D case.
   }\label{1}
    }
\end{center}
\end{figure}\vspace{0.5cm}

\hspace{-0.8 cm} We also see from figure-2 that for dust and
radiation cases $U(\phi)$ decreases with $\phi$ unlike the stiff
fluid case where it remains constant.\vspace{0.5cm}

\hspace{-0.8 cm} It may not be out of place to call attention to a
quintessential model driven by a tachyonic scalar
field~\cite{gorini} with a potential in 4D space time
\begin{equation}
V ( T ) = \frac{\Lambda}{\sin ^{2} \left( \frac{ 3 \sqrt{\Lambda
\gamma}}{2}\right) T}\sqrt{1 - \gamma \cos^{2}\left(\frac{ 3
\sqrt{\Lambda  \gamma}}{2}\right) T}
\end{equation}
($T$ is a tachyonic scalar field) giving the cosmological
evolution as
\begin{equation}
R (t) = R_{0} \left( \sinh \frac{3 \sqrt{\Lambda} \gamma
t}{2}\right)^{\frac{2}{3 \gamma}}
\end{equation}
It behaves like a two fluid model where one of the fluids is a
cosmological constant while the other obeys a state equation $ p =
(\gamma - 1) \rho $, $(0 < \gamma  < 2 )$.  Similarity of this
evolution with our model is more than apparent except for some
numerical factors coming out because we are  here dealing with a
higher dimensional spacetime. But the main result may be
re-emphasized that we get this evolution without forcing ourselves
to invoke any extraneous tachyonic type of scalar field.

\section*{3. Accelerating Universe - II }
As commented in the introduction the first model suffers from the
shortcoming that no dimensional reduction occurs. We here adopt
the conventional approach of taking a predetermined form of energy
momentum tensor and then try to solve the equations of motion with
the gauge condition
\begin{equation}
A(t) = R(t)^{-m}
\end{equation}\vspace{0.5cm}

\hspace{-0.6 cm}where $m$ is any positive number so that
dimensional reduction is ensured  \emph{a priori}. For the matter
field we here assume an equation of state of a generalised
Chaplygin type of gas in 3D space only

\begin{equation}
p=(\gamma -1)\rho - B \rho^{-\alpha}
\end{equation}
where $0\leq \alpha \leq 1$ along with a higher dimensional
isotropic pressure $p_{d}$ existing in the extra space  such the
the field equations reduce to
\begin{eqnarray}
\rho = \frac{k}{2} \frac{\dot{R}^{2}}{R^{2}}~~~~~~~~~~~~~~~~~~~~~~~~~~~~~~~~~~~~~~~~~~~~~~~~~~~\\
 -p =
(2-dm)\frac{\ddot{R}}{R} + \frac{1}{2}[m^{2}d( d+1) +
2(1-dm)]\frac{\dot{R}^{2}}{R^{2}}\\
-p_{d}= (3-dm +
m)\frac{\ddot{R}}{R}+\frac{1}{2}[m(d-1)(dm-4)+6]\frac{\dot{R}^{2}}{R^{2}}
\end{eqnarray}
where $k = dm^{2}(d-1) + 6(1-dm)$.\vspace{0.5cm}

\hspace{-0.6 cm}For positive energy density, $k$ must be greater
than zero which gives $m < \frac{3d-\sqrt{3d(d+2)}}{d(d-1)}$
 or, $m > \frac{3d+\sqrt{3d(d+2)}}{d(d-1)}$. \vspace{0.5cm}

\hspace{-0.6 cm}Again  Bianchi identity will be in our case
\begin{equation}
\dot{\rho} + 3\frac{\dot{R}}{R}(\rho + p) +
d\frac{\dot{A}}{A}(\rho + p_{d}) = 0
\end{equation}\vspace{0.5cm}

\hspace{-0.6 cm}Now using equations (36), (37) \& (40) we get

\begin{eqnarray}
\dot{\rho} + \frac{k}{(2-dm)}\frac{\dot{R}}{R}\left[\left\{\gamma
+ \frac{2 dm (m+1)}{k}\right\}\rho - B\rho^{-\alpha}\right] =0
\end{eqnarray}
Solving equation (41) we get

\begin{eqnarray}
\rho = \left[ \frac{Bk}{M} +
\frac{c}{R^{\frac{(1+\alpha)M}{(2-dm)}}}\right]^{\frac{1}{1+\alpha}}
\end{eqnarray}\vspace{0.5cm}

\hspace{-0.6 cm}where
\begin{equation}
M = \gamma k + 2dm (m+1)
\end{equation}
 and
$c$ is the integration constant. From physical considerations we
can finally put the restriction on $m$ as $m <
\frac{3d-\sqrt{3d(d+2)}}{d(d-1)}$ for $ d \neq 1$ and $m <1$ for $
d = 1$.\vspace{0.5cm}

\hspace{-0.6 cm}The relation (43) needs some interpretation. The
last term is a typical higher dimensional effect, absent in 4D
($d=0$). Now we see that several possibilities for different
values of $m$ present themselves.\vspace{0.5cm}

(\emph{i})  $m = - 1$:  Here $ A=R$ and we have to sacrifice the
desirable feature of dimensional reduction to get an isotropic
expansion in all dimensions. However it not exactly the cosmology
of the section I because we have a Chaplygin type of gas to start
with in this case. Moreover the isotropy of the metric dictates
that $p = p_{d}$ and so we end up with an isotropic pressure in
all dimensions. The solutions closely resemble the earlier work of
Debnath \emph{et al} ~\cite{ud} in 4D.\vspace{0.5cm}

 (\emph{ii})  $m = 0$:
 Here we get flat extra space although the total number of
 dimensions continues to be $(d+ 4)$. But the cosmology is exactly
 similar to the 4D case referred to earlier ~\cite{ud}. In fact
 this similarity is a direct consequence of a  little known theorem of Campbell
 that any analytic N-dimensional Riemmanian manifold can be
 locally embedded in a higher dimensional Ricci-flat manifold
 ~\cite{tavako}.\vspace{0.5cm}

(\emph{iii})  $d=0$:  Here we simply recover the 4D metric and all
the known solutions
  of 4D follow.\vspace{0.5cm}

\hspace{-0.8 cm}Now with the help of equations (37) \& (42), we
get

\begin{eqnarray}
\frac{\dot{R}^{2}}{R^{2}} = \frac{2}{k} \left[ \frac{Bk}{M} +
\frac{c}{R^{\frac{(1+\alpha)M}{(2-dm)}}}\right]^{\frac{1}{1+\alpha}}
\end{eqnarray}We have not been able, so far, to find a solution of equation (44)
in a closed form. Rather a Hypergeometric series solution results
given by

\begin{eqnarray}
2(2-dm)~  _{2}F_{1}\left[s,s,1+s,- \frac{B k \text{R}^{\frac{ M
}{2 s (2-dm)}}}{c M}\right]R^{\frac{M}{2(2-dm)}}
 = \frac{\sqrt{2} M c^{s}t }{\sqrt{k}}+C
\end{eqnarray}\vspace{0.5cm}

\hspace{-0.6 cm} where $s = \frac{1}{2(1+\alpha)}$ and $_{2}F_{1}$
is the hypergeometric function.\vspace{0.5cm}

\hspace{-0.6 cm}Even then, fixing the values of different
parameters one can get the temporal behaviour of the scale factors
as given in the adjoining figure-3. A cursory look at the figure
shows that at a certain stage of evolution the cosmology starts
inflating. Another desirable feature is the fact that the extra
dimensions compactify at very early stage of evolution in
conformity with both theoretical and observational requirements.

\begin{figure}[h]
\begin{center}

 \includegraphics[width=10 cm]{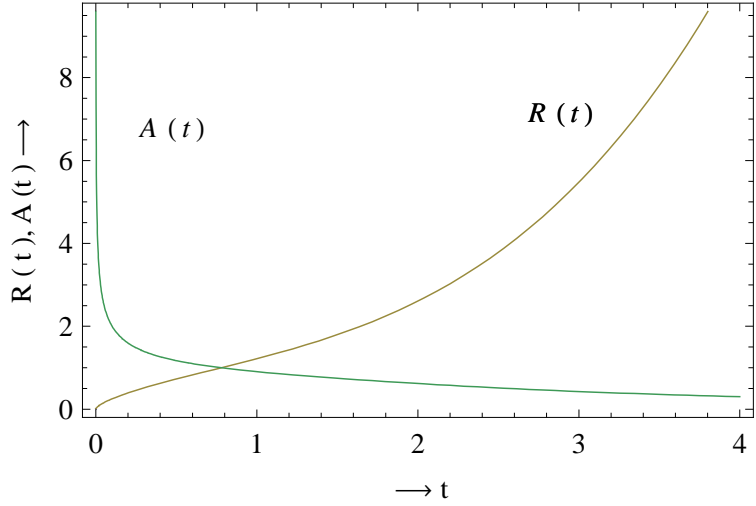}\\
  \caption{
  \small\emph{The variation of $R (t)$ and $A (t)$ vs $t$ is shown in this
  figure. Here we see that R (t) expands indefinitely and
  dimensional reduction is possible for extra dimensions. We have
  taken $\alpha$ = 0.5,  m = 0.5, d = 1 and $\gamma = 1$. This graph is drawn with the help of 'Mathematica'
  using equations (35) \& (44).}\label{1}
    }
\end{center}
\end{figure}\vspace{0.5cm}

\hspace{-0.6 cm}Now for small value of scale factor $R(t)$ ($A(t)$
should be large in this situation),
\begin{eqnarray}
\rho \approx c^{\frac{1}{1+\alpha}}R^{-\frac{M}{(2-dm)}}
\end{eqnarray}\vspace{0.1cm}

\hspace{-0.6 cm}which is very large and corresponds to the
universe dominated by an equation of state, $p = (\gamma - 1)\rho$
as is evident from equation (36). At the late stage of evolution
(when $R(t)$ is large and A(t) small) the expressions for $\rho$
and $p$ are given below.
\begin{eqnarray}
\rho \approx \left(\frac{Bk}{M}\right)^{\frac{1}{1+\alpha}}
                                                                \nonumber
 \mathrm{and}~ p \approx -
\left(\frac{Bk}{M}\right)^{\frac{1}{1+\alpha}} \left[\frac{M}{k} -
(\gamma -1)\right] = - \left[\frac{M}{k} - (\gamma
-1)\right]\rho\\
= - \left[ 1 + \frac{2dm(m+1)}{k} \right]\rho = \mathcal{W}\rho
\end{eqnarray}
\\where
\begin{eqnarray}
\mathcal{W} = - \left[ 1 + \frac{2dm(m+1)}{k} \right]
\end{eqnarray}\vspace{0.5cm}

\hspace{-0.6 cm}The last equation is interesting and introduces
some new physics in our analysis in the context of
multidimensional cosmology. Here $\mathcal{W} \neq -1 $. Obviously
this is due to the presence of extra dimensions in the above
equation. As mentioned in the introduction we have been working on
the idea of `\emph{dimension driven}' acceleration for the past
few years and the expression (47) points to the fact - how extra
dimensions aid the inflationary process.\vspace{0.5cm}

\hspace{-0.6 cm}In 4D case ($d = 0$) $\mathcal{W} = -1$ and a
$\Lambda$CDM model results. Otherwise the magnitude of
$\mathcal{W}$ is parameter dependent. When $ m = 0$, \emph{i.e.}
$A(t)$ is a constant we again get back the 4D case. This is,
however, a consequence of the Campbell's theorem~\cite{tavako} as
mentioned earlier.\vspace{0.5cm}

\hspace{-0.6 cm}When $m > 0$, $\mathcal{W} < -1$; So a phantom
like cosmology results with the occurrence of `big rip' etc. But
the cosmology becomes physically interesting when $-1<m<0$ such
that $0>\mathcal{W}>-1$ and we get a quiessence type of model
~\cite{hann}. In this connection one should note that predictions
from current measurements illustrate a degeneracy in which a flat
QCDM model in indistinguishable from closed $\Lambda$CDM model,
although the predicted value of $\mathcal{W}$ based on CMB
measurement is about $\mathcal{W} = - \frac{2}{3}$. However, the
error bars are too broad to reach firm conclusions.
\vspace{0.6cm}\\
It will be not out of space to call attention to a recent work by
Guo and Zhang et al\cite{gu} where a very generalised form of
Chaplygin relation is invoked where the constant B is assumed to
depend on the scale factor \emph{i.e.}, $ B = B(a)$. Taking $B =
B_{0}a^{-n}$, for example, it is shown that for a very large value
of the scale facor the model interpolates between a dust dominated
phase and a \emph{quiessence-dominated phase}(\emph{i.e.} , dark
energy with a constant equation of state)\cite{guo} given by $ W =
-1 + n/6 $. So our model closely mimics the above work where the
extra dimensions takes the role of a variable $B(a)$.
\hspace{-0.6 cm}\\
Now, if we calculate (3+d)-dimensional volume
\begin{eqnarray}
V = R^{3}A^{d}= R^{(3-dm)}
\end{eqnarray}
Since $m < \frac{3d-\sqrt{3d(d+2)}}{d(d-1)}$ for $ d \neq 1$ and
$m <1$ for $ d = 1$, $\therefore$ the value of $(3 - dm)$ should
be positive.
\\Now for accelerating universe, $\ddot{R} > 0$, implying
\begin{eqnarray}
R^{\frac{M(1+\alpha)}{(2-dm)}} >
\frac{c}{2B}\frac{M(M+2dm-4)}{k(2-dm)}
\end{eqnarray}
 \\The above expression shows that the universe will be decelerating
for small values of scale factor while for large values  we get
accelerating universe and the flip occurs at $t= t_{c}$ such that

\begin{eqnarray}
R(t_{c}) =
\left[\frac{c}{2B}\frac{M(M+2dm-4)}{k(2-dm)}\right]^\frac{2-dm}{M(1+\alpha)}
\end{eqnarray}
\begin{figure}[h]
\begin{center}
  \includegraphics[width=10 cm]{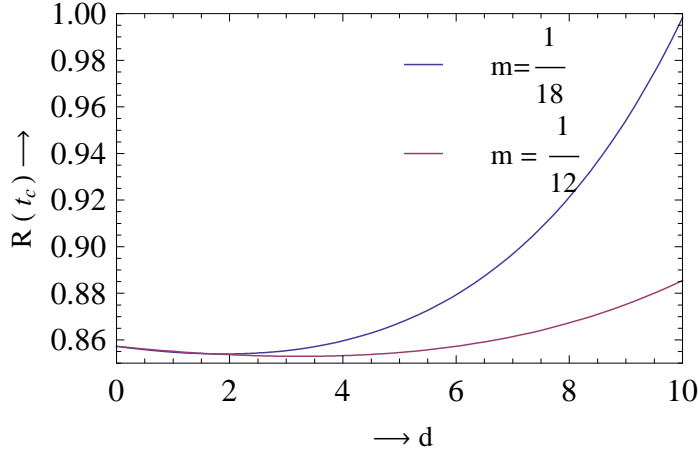}\\
  \caption{
  \small\emph{The variation of $R (t_{c})$ and $d$ is shown in this figure for different values of $m$.
  Here we have taken $\alpha$ = 0.5 and $\gamma = 1$. }\label{1}
    }
\end{center}
\end{figure}\vspace{0.5cm}

\hspace{-0.8 cm} While looking at the plot (figure-4) one must
admit that it is rather artificial to take the number of spatial
dimensions as a continuous variable. However given the fact that
in Friedmannian cosmology $R(t)$ is always a monotonically
increasing function of time, $R(t)$ may be taken as a measure of
the flipping instant and incidentally it does not initially show
much variation with $d$. The fact that it shoots up at some higher
value of $d$ need not be taken too seriously because it may be due
to the arbitrarily chosen initial value of $\alpha$, $m$ etc. In
any case the \emph{flip} is delayed with dimensions and from
observational point it is encouraging because all the evidences
from
 different cosmic probes point to a very late acceleration.\vspace{0.5cm}

\hspace{-0.6 cm} The speed of sound $v_{s} = \frac{\partial
p}{\partial \rho}$ in the Chaplygin gas

\begin{eqnarray}
v_{s}^{2} = (\gamma - 1)(1 + \alpha) - \frac{\alpha p}{\rho}
\end{eqnarray}
\\Using equations (38), (39) $\&$ (52) we get,

\begin{eqnarray}
v_{s}^{2} = \frac{1 + \alpha}{d + 3} +
\frac{\alpha}{k}\left\{m^{2}d(d + 1) + 2(1 - dm)\right\} +
\frac{2\alpha q}{k}(2 - dm)
\end{eqnarray}\vspace{0.5cm}

\hspace{-0.6 cm}where $q $ is the deceleration parameter defined
as $ - \frac{\ddot{R}R}{\dot{R}^{2}}$. For accelerating model the
value of $q$ should be negative. So for real value of $v_{s}$,
$\frac{1 + \alpha}{d + 3} + \frac{\alpha}{k}\left\{m^{2}d(d + 1) +
2(1 - dm)\right\} > \frac{2\alpha \mid q \mid}{k}(2 - dm)$ which
gives $\frac{k(1 + \alpha)}{2 \alpha (d + 3)(2 - dm)} +
\frac{1}{2(2-dm)}\left\{m^{2}d(d + 1) + 2(1-dm)\right\} > \mid q
\mid$. This relation gives an upperbound for the $q$, which may
have observational consequences. Since $0<\alpha<1$, it can not
exceed the speed of light. But for larger values of $\alpha$ the
speed of sound exceeds the speed of light. It also depends upon
the value of $m$. In cosmology a speed of sound exceeding the
speed of light does not contradict the causality \cite{Gorini}.
 \\Let $\alpha = 0.5$, $m = \frac{1}{6}$, and $d = 6$, therefore $k$
should be $\frac{5}{6}$ and $M$ equal to $ \frac{88}{27}$ in this
case. If we calculate $v_{s}^{2}$, we get two conditions (i) when
$\mid q \mid < 0.923 $ then $v_{s} < c$, (ii) otherwise $v_{s} > c$.
 \\Following the analysis given in the previous section we get the
analogous energy density $\rho_{\phi}$ and pressure $p_{\phi}$ for
the scalar field as

\begin{eqnarray}
\rho_{\phi}=\frac{1}{2}\dot{\phi}^{2}+ U(\phi) = \left[
\frac{Bk}{M} +
\frac{c}{R^{\frac{(1+\alpha)M}{(2-dm)}}}\right]^{\frac{1}{1+\alpha}}~~~~~~~~~~~~~~~
\end{eqnarray}
\begin{eqnarray}
 p_{\phi} =\frac{1}{2}\dot{\phi}^{2}- U(\phi) = (\gamma -1)\rho - B
 \rho^{-\alpha}~~~~~~~~~~~~~~~~~~~~~~~~~~~~~~ ~~~~~~~~~~~~~~~\\
~~~~~~~~~~~~~~~~~~~~ = (\gamma -1)\left[\frac{Bk}{M}
 +\frac{c}{R^{\frac{(1+\alpha)M}{(2-dm)}}}\right]^{\frac{1}{1+\alpha}}-B \left[\frac{Bk}{M}
 +\frac{c}{R^{\frac{(1+\alpha)M}{(2-dm)}}}\right]^{\frac{\alpha}{1+\alpha}}
\end{eqnarray}\vspace{0.5cm}

\hspace{-0.6 cm}such that we get
\begin{eqnarray}
 \dot{\phi}^{2} = \left[\frac{Bk}{M}
 +\frac{c}{R^{\frac{(1+\alpha)M}{(2-dm)}}}\right]^{-\frac{\alpha}{1+\alpha}} \left[\left( \gamma
 \frac{Bk}{M}-B \right)
 +\frac{c \gamma}{R^{\frac{(1+\alpha)M}{(2-dm)}}}\right]
\end{eqnarray}
\begin{eqnarray}
 \phi' = \pm \frac{1}{R}\sqrt{\frac{k}{2}}\left[\frac{Bk}{M}
 +\frac{c}{R^{\frac{(1+\alpha)M}{(2-dm)}}}\right]^{-\frac{1}{2}} \left[B \left( \gamma
 \frac{k}{M}- 1 \right)
 +\frac{c \gamma}{R^{\frac{(1+\alpha)M}{(2-dm)}}}\right]^{\frac{1}{2}}
\end{eqnarray}\vspace{0.5cm}

\hspace{-0.6 cm}and
 \begin{eqnarray}
 U(\phi) =
\frac{1}{2}(2- \gamma)\left[\frac{Bk}{M}
 +\frac{c}{R^{\frac{(1+\alpha)M}{(2-dm)}}}\right]^{-\frac{1}{1+\alpha}} + \frac{B}{2}\left[  \frac{Bk}{M}
 +\frac{c}{R^{\frac{(1+\alpha)M}{(2-dm)}}}\right]^{-\frac{\alpha}{1+\alpha}}
\end{eqnarray}\vspace{0.5cm}

\hspace{-0.6 cm} Integrating equation (58)

\begin{eqnarray}
 \phi -  \phi _{0} = \pm \int \left[\frac{1}{R}\sqrt{\frac{k}{2}}\left\{\frac{ B \left(\gamma
 \frac{k}{M}- 1 \right)
 +\frac{c \gamma}{R^{\frac{(1+\alpha)M}{(2-dm)}}}}{\frac{Bk}{M}
 +\frac{c}{R^{\frac{(1+\alpha)M}{(2-dm)}}}}\right\}^{\frac{1}{2}} \right]dR
\end{eqnarray}\vspace{0.5cm}

\hspace{-0.6 cm}where $\phi _{0}$ is an arbitrary constant of
integration. It is very difficult to integrate the equation (60)
in a closed form. However if at this stage  we take the $m = -1$
\emph{i.e.}, $\frac{\gamma k}{M} = 1$ then we are able to
integrate and get the following solution.

\begin{eqnarray}
 \phi  = \sqrt{\frac{2}{k \gamma}}\frac{2+d}{1+\alpha}\sinh^{-1}\left\{\sqrt{\frac{c
 \gamma}{B}}R^{-\frac{(1+\alpha) k \gamma}{2(2+d)}}\right\}
\end{eqnarray}
and
\begin{figure}[t]
\begin{center}
 \includegraphics[width=8 cm]{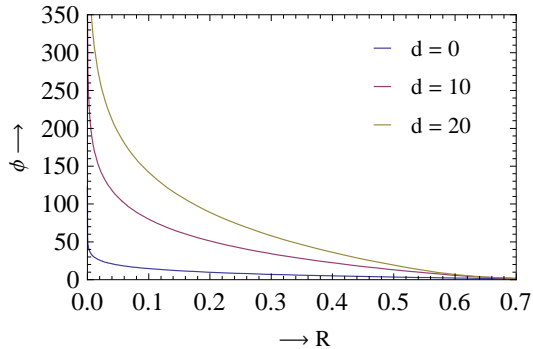}\\
  \caption{
  \small\emph{The variation of $\phi$ and $R$ is shown in this
  figure. The scalar field decays with time more sharply in lower
  dimensions. (Taking $\alpha$ = 0.6,  k = 6 and $\gamma = 1$).}\label{1}
    }
\end{center}
\end{figure}
\begin{eqnarray}
 U(\phi ) = \frac{2-\gamma}{2}\left(\frac{B}{\gamma}\right)^{\frac{1}{1+
 \alpha}}\cosh^{\frac{2}{1+\alpha}}\left\{\sqrt{\frac{k \gamma}{2}}\frac{1 + \alpha}{d +
 2}\phi\right\}\\ \nonumber
 + ~\frac{B}{2}\left(\frac{B}{\gamma}\right)^{-\frac{\alpha}{1+
 \alpha}}\cosh^{-\frac{2 \alpha}{1+\alpha}}\left\{\sqrt{\frac{k \gamma}{2}}\frac{1 + \alpha}{d +
 2}\phi\right\}
\end{eqnarray}
\begin{figure}[h]
\begin{center}
 \includegraphics[width=7.8 cm]{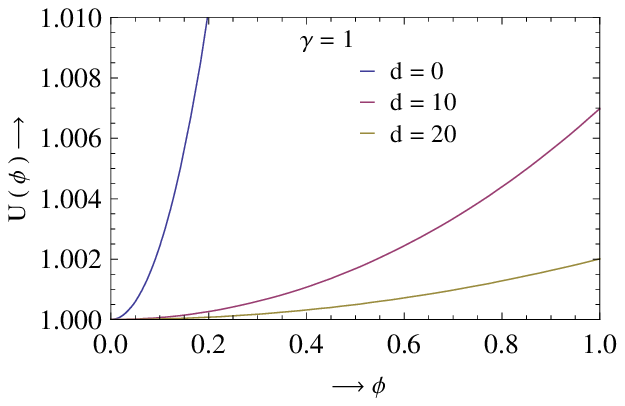}
 \includegraphics[width=7.8 cm]{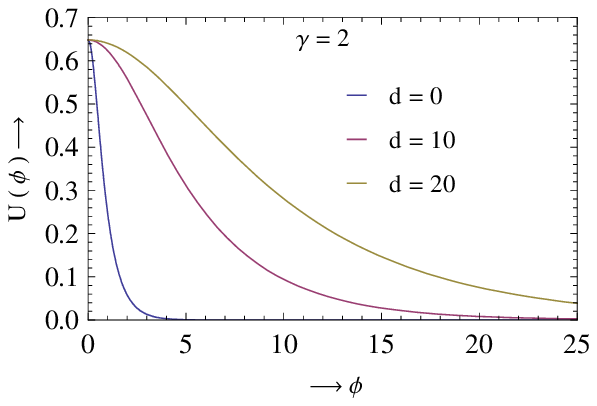}\\
  \caption{
  \small\emph{The variation of $U (\phi) $ and $\phi$ is
  shown. It suggests critical $\gamma$ dependance. With dimensions $U (\phi) $ becomes flatter.
 ($\alpha$ = 0.6, k = 6).  }\label{1}
    }
\end{center}
\end{figure}
 If we put $d = 0$, \emph{i.e.}, in 4-Dimensional spacetime we get
solutions of the above equations (61) $\&$ (62) which are
identical with 4D case ~\cite{ud}.\vspace{0.5cm}

\section*{4. Discussion }
We have here studied two higher dimensional cosmological models
amenable to late acceleration - one with a particular form of
deceleration parameter and for the other with a generalised
chaplygin type of gas. The important findings may be briefly
summarised as:\vspace{0.5cm}

(\emph{i}) Unlike the first case we have dimensional reduction in
the second.  \vspace{0.5cm}

 (\emph{ii}) The striking thing about the first
model is the fact that we do not have to assume an extraneous,
unphysical
 quintessential type of matter field \emph{a priori } to
achieve the late acceleration.   \vspace{0.5cm}

 (\emph{iii}) It is observed that the instant of
\emph{flip} is delayed with dimensions, which is also supported by
current observations. In this respect the HD accelerating models
have an edge over the 4D ones.  \vspace{0.5cm}

 (\emph{iv}) The most important thing in the
second case, in our opinion, is the finding that depending on some
initial conditions, the effective equation of state during late
evolution interpolates among $\Lambda CDM$, QCDM and Phantom type
of expansion. In this respect our work recovers the effective
equation of state (for large scale factor) for a recent work of
Guo et al where a very generalised Chaplygin type of gas is
taken.\vspace{0.5cm}

(\emph{v}) Our solutions are general in nature because all the
known results of 4D cosmology are recovered when $d = 0$.

\textbf{Acknowledgment : }SC acknowledges the financial support of
UGC, New Delhi for a Major Research Project award while DP
acknowledges financial support of ERO, UGC  for a Minor Research
project.


\begin{thebibliography}{40}
\bibitem{pd} T. Padmanabhan  - `Understanding our
Universe : Current status and open issues'  and references
therein, gr- qc/0503107
\end{thebibliography}
\end{document}